\documentclass[10pt,prl,aps,twocolumn,superscriptaddress,floatfix,showpacs]{revtex4}

\pdfoutput=1

\usepackage{graphicx}
\usepackage{amssymb}
\usepackage{bm}
\usepackage{bbold}
\usepackage{color}

\begin{document}

\title{A tensor product state approach to spin-1/2 square $J_1$-$J_2$ antiferromagnetic Heisenberg model: evidence for deconfined quantum criticality}

\author{Ling Wang}
\affiliation{Beijing Computational Science Research Center, 10 West Dongbeiwang Road, Beijing 100193, China}
\affiliation{Institute of Quantum Information and Matter and Department of Physics, California Institute of Technology, Pasadena, CA 91125, USA}

\author{Zheng-Cheng Gu}
   \affiliation{ Department of Physics, The Chinese University of Hong Kong, Shatin, New Territories, Hong Kong}
\affiliation{Perimeter Institute for Theoretical Physics, Waterloo, Ontario, N2L2Y5, Canada}

\author{Frank Verstraete}
\affiliation{Vienna Center for Quantum Science and Technology, Faculty of Physics,
University of Vienna, Boltzmanngasse 5, 1090 Vienna, Austria}
\affiliation{Department of Physics and Astronomy, Ghent University, Krijgslaan 281-S9, B-9000 Gent, Belgium}

\author{Xiao-Gang Wen}
\affiliation{Department of Physics, Massachusetts Institute of Technology, Cambridge, Massachusetts 02139, USA}
\affiliation{Perimeter Institute for Theoretical Physics, Waterloo, Ontario, N2L2Y5, Canada}

\date{\today}

\begin{abstract}
  The ground state phase of spin-1/2 $J_1$-$J_2$ antiferromagnetic
  Heisenberg model on square lattice around the maximally frustrated
  regime ($J_2\sim 0.5J_1$) has been debated for decades.  Here we
  study this model using the cluster update algorithm for tensor
  product states (TPSs). The ground state energies at finite sizes and
  in the thermodynamic limit (with finite size scaling) are in good
  agreement with exact diagonalization study. Through finite size
  scaling of the spin correlation function, we find the critical point
  $J_2^{c_1}=0.572(5)J_1$ and critical exponents $\nu=0.50(8)$,
  $\eta_s=0.28(6)$. In the range of $0.572 < J_2/J_1 \leqslant
  0.6 $ we find a paramagnetic ground state with exponentially
  decaying spin-spin correlation. Up to $24\times 24$ system size, we
  observe power law decaying dimer-dimer and plaquette-plaquette
  correlations with an anomalous plaquette scaling exponent
  $\eta_p=0.24(1)$ and an anomalous columnar scaling exponent
  $\eta_c=0.28(1)$ at $J_2/J_1=0.6$. These results are consistent with
  a potential gapless $U(1)$ spin liquid phase. However, since the
  $U(1)$ spin liquid is unstable due to the instanton effect, a VBS
  order with very small amplitude might develop in the thermodynamic
  limit. Thus, our numerical results strongly indicate a deconfined
  quantum critical point (DQCP) at $J_2^{c_1}$.  Remarkably, all the
  observed critical exponents are consistent with the $J-Q$ model.
  
\end{abstract}

\maketitle

\section{Introduction}
The spin 1/2 $J_1$-$J_2$ antiferromagnetic Heisenberg model on a
square lattice has drawn great attention for the last two decades
owing to its close relation to the disappearance of the
antiferromagnetic (AF) long range order (LRO) in the high-$T_c$
superconducting materials~\cite{RVB,PatrickRVB}, and has been proposed
as a possible simple model to realize topologically ordered chiral
spin-liquid state~\cite{KL8795,WWZ8913} or $Z_2$ spin-liquid
state~\cite{RS9173,W9164,SondhiZ2,HongZ2,HuJ1J2}. The Hamiltonian of this
model is given by:
\begin{equation}
\label{model}
H=J_1\sum_{(i,j)}\mathbf{S_i}\cdot\mathbf{S_j}+J_2\sum_{\langle
i,j\rangle}\mathbf{S_i}\cdot\mathbf{S_j},\quad
(J_1,J_2>0),
\end{equation}
where $(i,j)$ represents the nearest neighbor (NN) pair and $\langle
i,j\rangle$ represents the next nearest neighbor (NNN) pair. For
convenience, we set $J_1=1$ throughout the paper. It has long been
believed that the frustration from NNN interaction competes with the
NN one and drives the system through a quantum phase transition from
an AF LRO phase to a magnetically disordered phase. In two extreme
cases, the ground state phases of the model are well established: at
very small $J_2$, the ground state has AF LRO; and at very large
$J_2$, the system falls into two weakly coupled sets, and the magnetic
susceptibility peaks at momentum $(\pi,0)$ or $(0,\pi)$. In the
intermediate coupling regime, quantum fluctuation is meant to destroy
the AF LRO near the maximally frustrated point $J_2=0.5$ of the
classical model and establishes a new paramagnetic phase. The nature
of such a quantum phase is of great interest.

Numerous efforts have been made using many different approaches, such
as the exact diagonalization
(ED)~\cite{ed1,ed2,ed3,ed4,ed5,richter_ed}, spin-wave
theory~\cite{chandra_prb38.9335,ivanov_prb46.8206}, series
expansion~\cite{arlego_prb78.224415,sirker_prb73.184420}, large-N
expansion~\cite{read_prl66.1773}, the coupled cluster method
(CCM)~\cite{darradi_prb78.214415}, variational methods (including
short range resonating valence bond (SRVB)
method)~\cite{kevin_prb79.224431,mambrini_prb74.144422,Capriotti_prl87.097201},
and the fixed-node quantum monte carlo
(QMC)~\cite{Capriotti_prl84.3173}. The results turned out being
controversial: a series expansion calculation of a general magnetic
susceptibility over different perturbation fields suggests that within
the Ginzburg-Landau paradigm the type of phase transition from the
N\'eel to paramagnetic phase is of first
order~\cite{sirker_prb73.184420}. However, the same general magnetic
susceptibility calculated with a coupled cluster method suggests a
second order phase transition~\cite{darradi_prb78.214415}. The nature
of the phase near $J_2=0.5$ was as unclear: a fixed-node QMC study
indicates a plaquette valence bond solid (VBS)
state~\cite{Capriotti_prl84.3173}; whereas the series expansion argues
for a columnar VBS state~\cite{sirker_prb73.184420}. A relatively
direct investigation of the nature of the ground state order is using
the SRVB approximation~\cite{mambrini_prb74.144422}, where with
another term $J_3$ included in the Hamiltonian, a plaquette VBS state
along the line of $J_2+J_3=0.5$ is found. Most recently, density
matrix renormalization group (DMRG) has demonstrated its power in
simulating quasi-one-dimensional cylinders for the Kagome Heisenberg
model. Different groups applied it to the spin $1/2$ $J_1-J_2$ model
as well, however the results were different: Jiang et al. claim a
$Z_2$ spin liquid state~\cite{JiangJ1J2} while Gong et al. suggests a
plaquette VBS state~\cite{ShengJ1J2}. 

\begin{figure}
\begin{center}
\includegraphics[width=8cm]{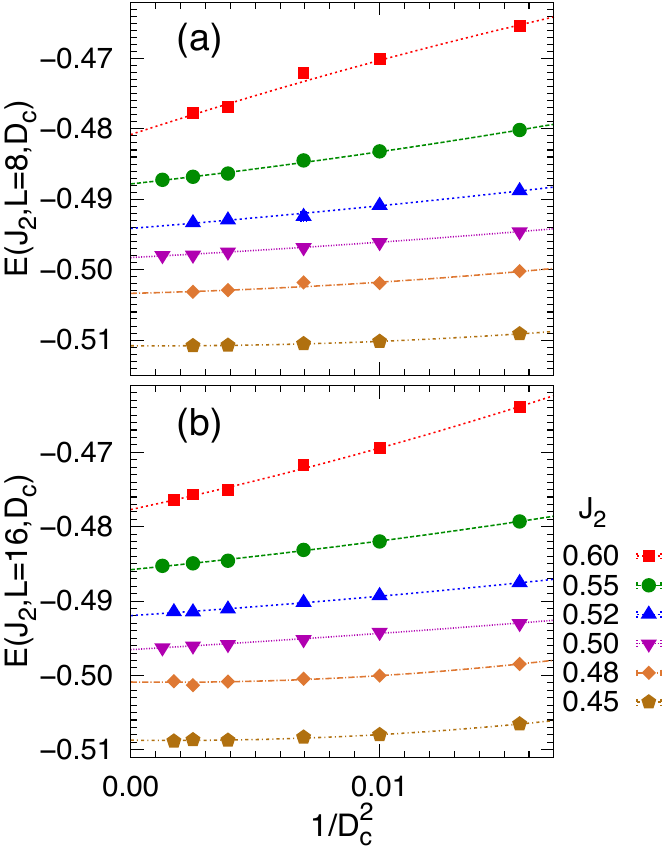}
\caption{The finite size ground state energies using the largest
  available bond dimension $D=9$ measured at various
  $D_c=8,10,12,16,20,24,28$ in a VMC-tensor renormalization
  algorithm~\cite{ling}. The finite size energies are extrapolated to
  $D_c\to \infty$ limit by fitting to second order polynomials, the
  fitted results are in dashed lines.}
\label{enrplot}
\end{center}
\end{figure}

In this paper, we revisit this problem with a TPS~\cite{frank04}
ansatz for the ground state wave function, accessed by the recently
proposed cluster update algorithm~\cite{ling_cluster}, and reveal the
answer to both questions. By fitting a universal scaling function for
the spin-spin correlation we observe a continuous phase transition
from the N\'eel to paramagnetic phase at $J_2^{c_1} = 0.572(5)$ with
critical exponents $\nu=0.50(8)$ and $\eta_s=0.28(6)$. In the
paramagnetic phase we find an exponentially decaying spin-spin
correlation functions. Up to $24\times 24$ system size, we observe
power law decaying dimer-dimer and plaquette-plaquette correlation
functions, which indicate a non-zero spin triplet gap and zero spin
singlet gap. These properties are consistent with the previously
proposed $U(1)$ gapless spin liquid state~\cite{WangJ1J2} by using a
one-parameter TPS ansatz. Nevertheless, it is well known that the
$U(1)$ gapless spin liquid is unstable due to instanton effect, we
argue that a very small VBS order might eventually develop in the
thermodynamic limit. Interestingly, at $J_2=0.6$ the scaling of
plaquette (columnar) VBS order parameter suggests an anomalous VBS
scaling exponent $\eta_p=0.24(1)$ ($\eta_c=0.28(1)$). Remarkably, all
observed exponents are consistent with that of the $J-Q$
model~\cite{DQCP3,DQCP4,DQCP6}. Thus, our numerical results strongly
indicate a deconfined quantum critical point scenario
(DQCP)~\cite{DQCP1,DQCP2} from Neel order to VBS order at
$J_2=0.572(5)$.

\begin{figure}
\begin{center}
\includegraphics[width=8cm]{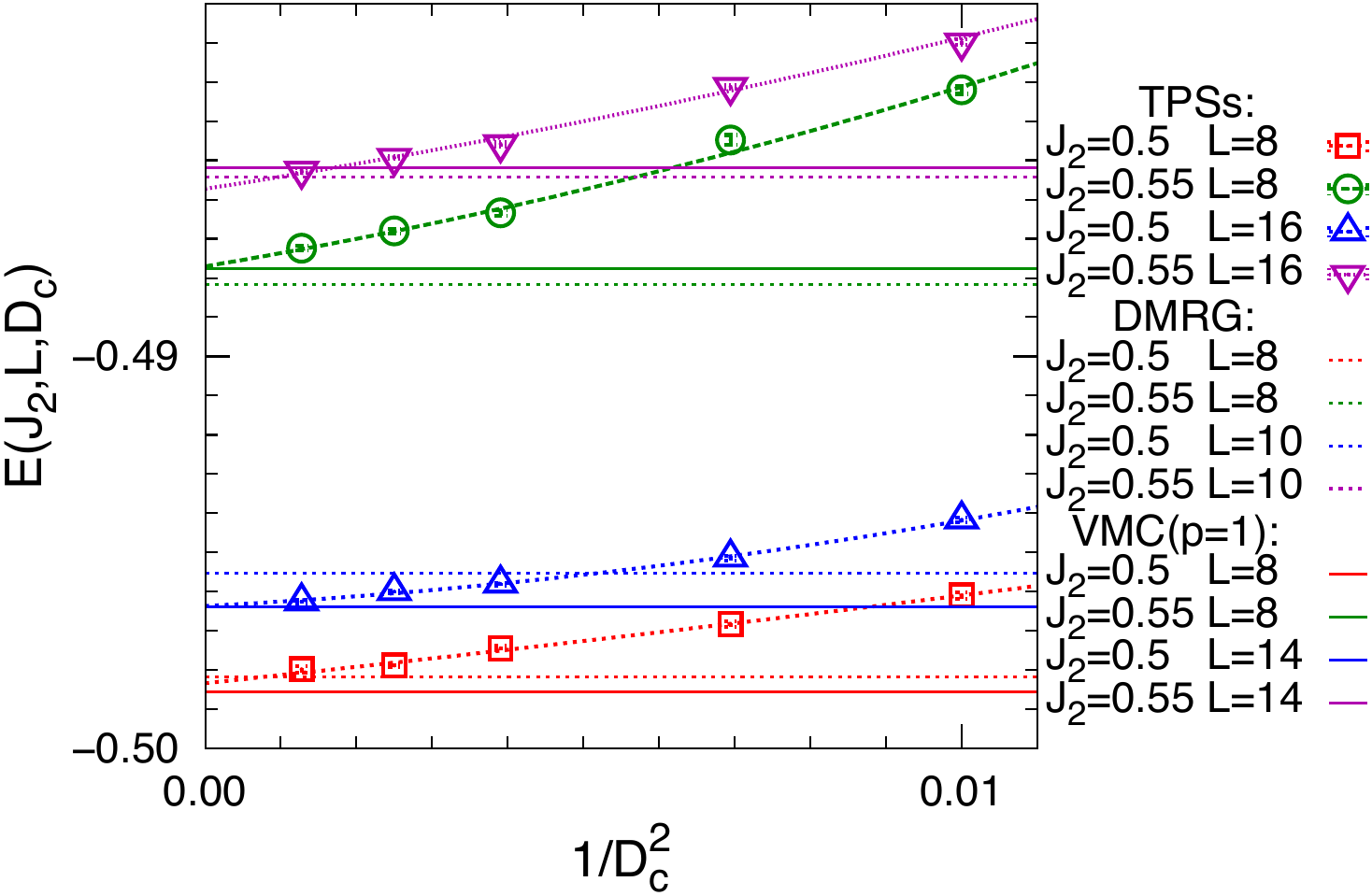}
\caption{A benchmark of ground state energy with the SU(2) symmetric
  DMRG results~\cite{ShengJ1J2} on tori and the VMC calculation with
  one Lanczos projection step~\cite{HuJ1J2} on tori at $J_2=0.5,0.55$,
  where $D_c$ is the Schmidt number kept in our VMC-tensor
  renormalization algorithm.}
\label{compareDMRG}
\end{center}
\end{figure}

\section{Results}
We divide the square lattice into four sublattices $A,B,C,D$ that form
a $2\times 2$ unit cell, and associate each type of sites with one of
the four different sublattice tensors. Such a choice of tensor product
state ansatz aims at describing potential VBS orders and studying
their competing effects. We use the cluster update imaginary time
evolution method~\cite{ling_cluster} to evolve from a TPS with random
initialized tensor elements to a converged state of the $J_1-J_2$
Hamiltonian. Such obtained state is often called infinite TPS, since
there is no system size information entering into this evolution
scheme. Once the infinite TPS (with a bond dimension $D=9$) converges,
we cover a $L\times L$ torus with repeated $2\times 2$ unit cells and
form a finite size wavefunction. Without further finite size ground
state optimization, we measure the size dependent variational
energies, staggered magnetizations, spin-spin, dimer-dimer and
plaquette-plaquette correlations on $L\times L$ tori for
$L=4,6,8,12,16,24$. Our tensor measurement strategy is to use
variational monte carlo (VMC)~\cite{ling} to sample spin
configurations, whose weights are computed by the tensor
renormalization idea~\cite{Levin07,gu2008}. With these combined
techniques, we can make precise measurements (for a periodic systems)
and perform careful finite size scaling analysis. 
Here and after, all
our numerical results are obtained from TPSs with bond dimension
$D=9$.

\begin{figure}
\begin{center}
\includegraphics[width=7cm]{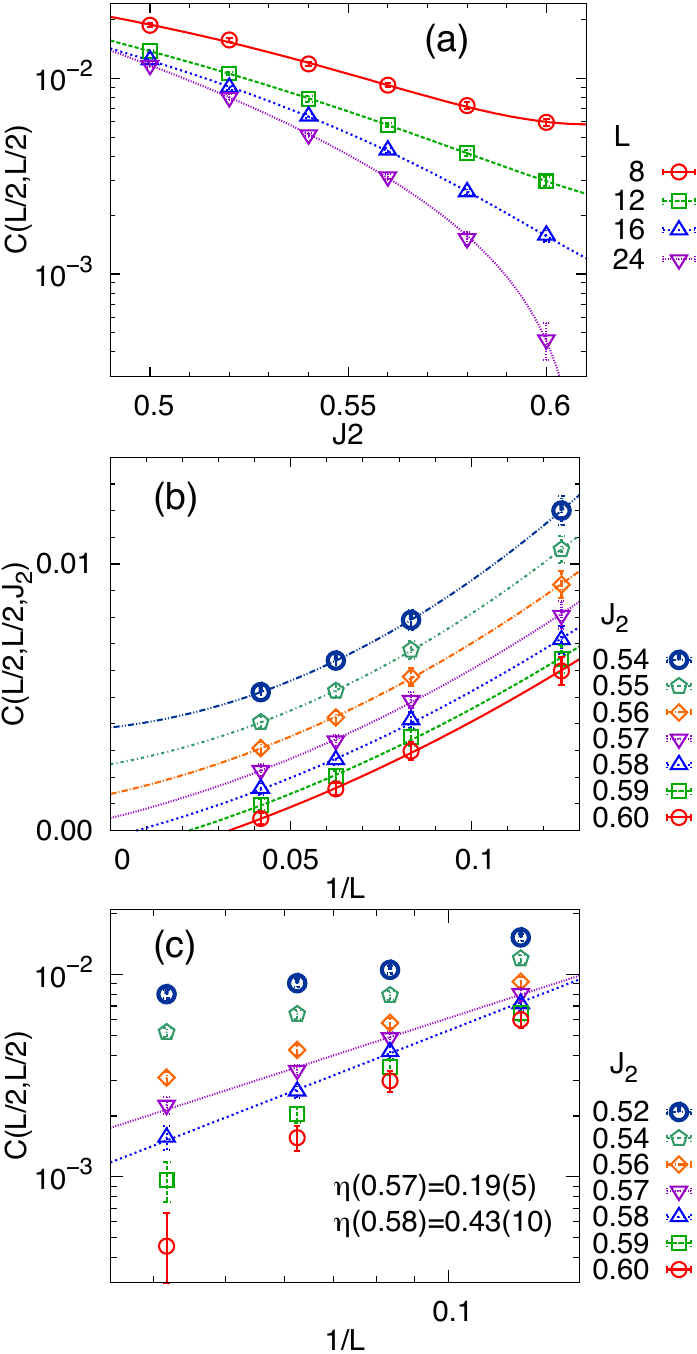}
\caption{(a) The largest distance spin-spin correlation as a function
  of $J_2$ at $L=8,12,16,24$. The same correlations $C(L/2,L/2)$
  presented against $1/L$ in a regular plot (b) and in a {\it log-log}
  plot (c) for various $J_2$.}
\label{magplot}
\end{center}
\end{figure}
 
\emph{Ground state energies} -- We present the ground state energies
on tori of $L\times L$ (at $L=8,16$) as functions of $D_c$
($D_c=8,10,12,16,20,24,28$) in Fig.~\ref{enrplot}, where $D_c$ is the
Schmidt numbers kept in calculating the importance weight of sampled
spin configurations~\cite{ling}. We find that the variational ground
state energies decrease monotonically with increasing $D_c$. Using a
quadratic function in $1/D_c^2$ we extrapolate the finite size
energies to the $D_c\to \infty$ limit. The fitted results are shown in
dashed curves in Fig.~\ref{enrplot}. Note that our measurement
scheme~\cite{ling} makes approximations to the importance weight of
sampled spin configurations, where error is controlled by $D_c$.
However the VMC principle guarantees that all measured energies at
finite $D_c$ are variational. Furthermore, the almost-converged
energies at $D_c=28$ make sure that the extrapolations are reliable.

We compare our variational energies on tori with the $SU(2)$ symmetric
DMRG results on tori~\cite{ShengJ1J2} and the best VMC with Lanczos
projection steps~\cite{HuJ1J2} on tori. As seen in
Fig.~\ref{compareDMRG}, at system size $L=8$, our results are
consistent with the DMRG results. Interestingly, our variational
energies on tori of $L=16$ are lower than the DMRG energies on tori of
$L=10$. This means that the ground states entanglement on tori of
$10\times 10$ are beyond the resolution of the DMRG if keeping only
8000 $SU(2)$ Schmidt states, which also explains why DMRG often
relies on cylinder studies instead of tori. Very impressively, our
variational ground state energy for $L=16$ is comparable to (at
$J_2=0.5$) or better than (at $J_2=0.55$) the best VMC results for a
smaller size $L=14$ torus with one steps Lanczos
projections~\cite{HuJ1J2}.

\begin{figure}
\begin{center}
\includegraphics[width=7cm]{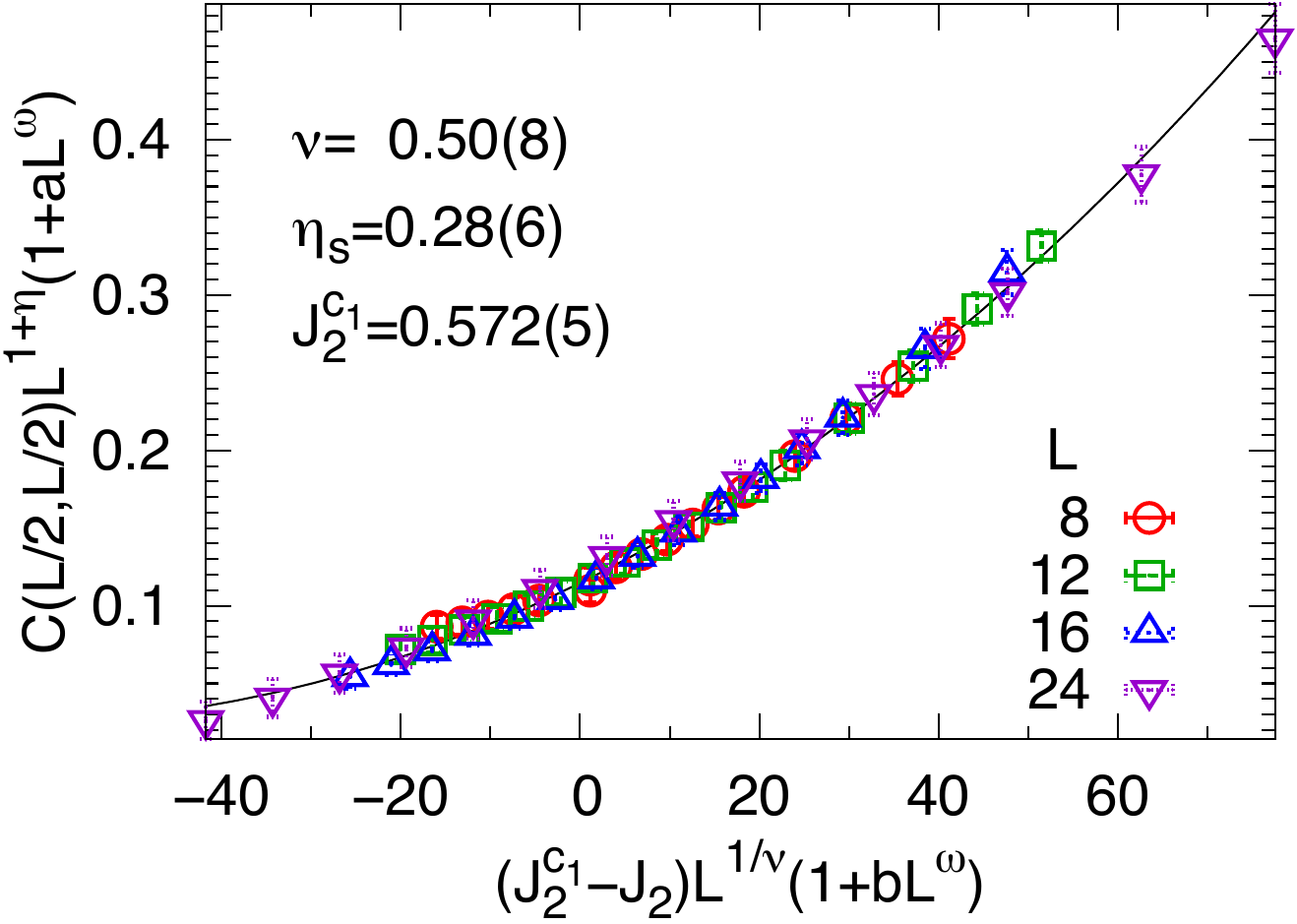}
\caption{The finite size scaling function of $C(L/2,L/2)$.}
\label{fss}
\end{center}
\end{figure}

\emph{Staggered magnetization} --
The staggered magnetization square is defined as 
\begin{equation}
M^2=\frac{1}{N}\sum_{r_x,r_y}(-1)^{r_x+r_y}C(r_x,r_y),
\end{equation}
where $C(r_x,r_y)$ is the spin-spin correlation function
\begin{equation}
C(r_x,r_y)=\frac{1}{N}\sum_{x,y}\mathbf{S}_{(x,y)}\cdot\mathbf{S}_{(x+r_x,y+r_y)}.
\end{equation}
We compute the spin-spin correlation functions at the largest distance
$C(L/2,L/2)$ for various $L$, and show their dependence with coupling
$J_2$ in Fig.~\ref{magplot}(a), here dashed lines are polynomial
fittings. To determine the critical transition point, we present
$C(L/2,L/2)$ against $1/L$ in a regular plot for various $J_2$ in
Fig.~\ref{magplot}(b), and extrapolate using quadratic functions
(shown in dashed lines). We find the critical point to be
$0.57<J_2^{c_1}<0.58$. To see the critical behavior, we present
$C(L/2,L/2)$ versus $1/L$ in a {\it log-log} plot in
Fig.~\ref{magplot}(c). Taking the critical scaling as
$C(L/2,L/2)\varpropto L^{-(z+\eta_s)}$ and using a linear regression
function, we find the anomalous spin scaling exponent $\eta_s=0.19(5)$
at $J_2=0.57$ and $\eta_s=0.43(10)$ at $J_2=0.58$. Finally, We take
the finite size scaling (FSS) formula
\begin{equation}
  C(L/2,L/2)L^{z+\eta_s}(1+aL^{\omega})=F\left((J^{c_1}_2-J_2)L^{1/\nu}(1+bL^{\omega})\right)
\end{equation}
to determine the critical point $J_2^{c_1}=0.572(5)$ and the critical
exponents $\nu=0.50(8)$, $\eta_s=0.28(6)$, with result presented in
Fig.~\ref{fss}. Here $F(x)$ is a dimensionless polynomial, $\omega$
represents sub-leading finite size corrections whose value is set to
2.

\begin{figure}
\begin{center}
\includegraphics[width=7cm]{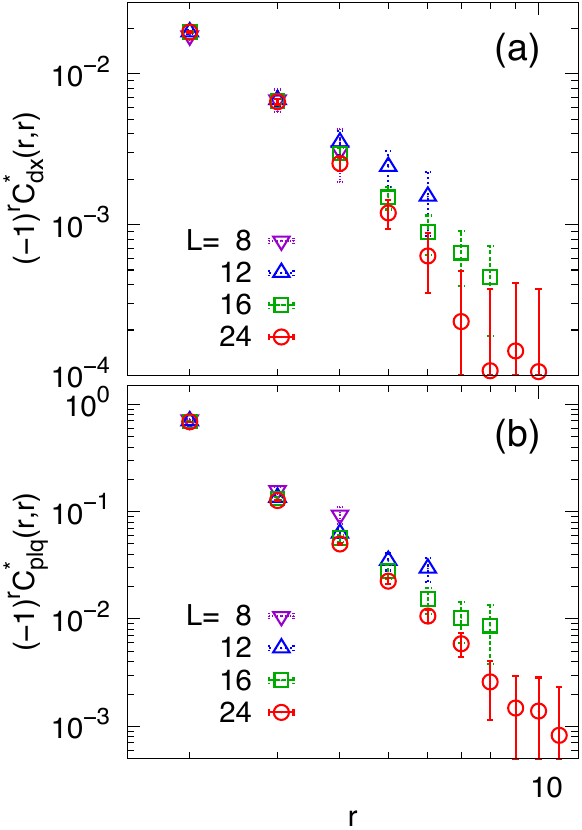}
\caption{ The modified dimer-dimer correlation $C^*_{dx}(r,r)$ (a)
  and plaquette-plaquette correlation $C^*_{plq}(r,r)$ (b) as a function
  of separation $r$ at $J_2=0.6$ in {\it log-log} plots.}
\label{correlationfuncs}
\end{center}
\end{figure}

\begin{figure}
\begin{center}
\includegraphics[width=7cm]{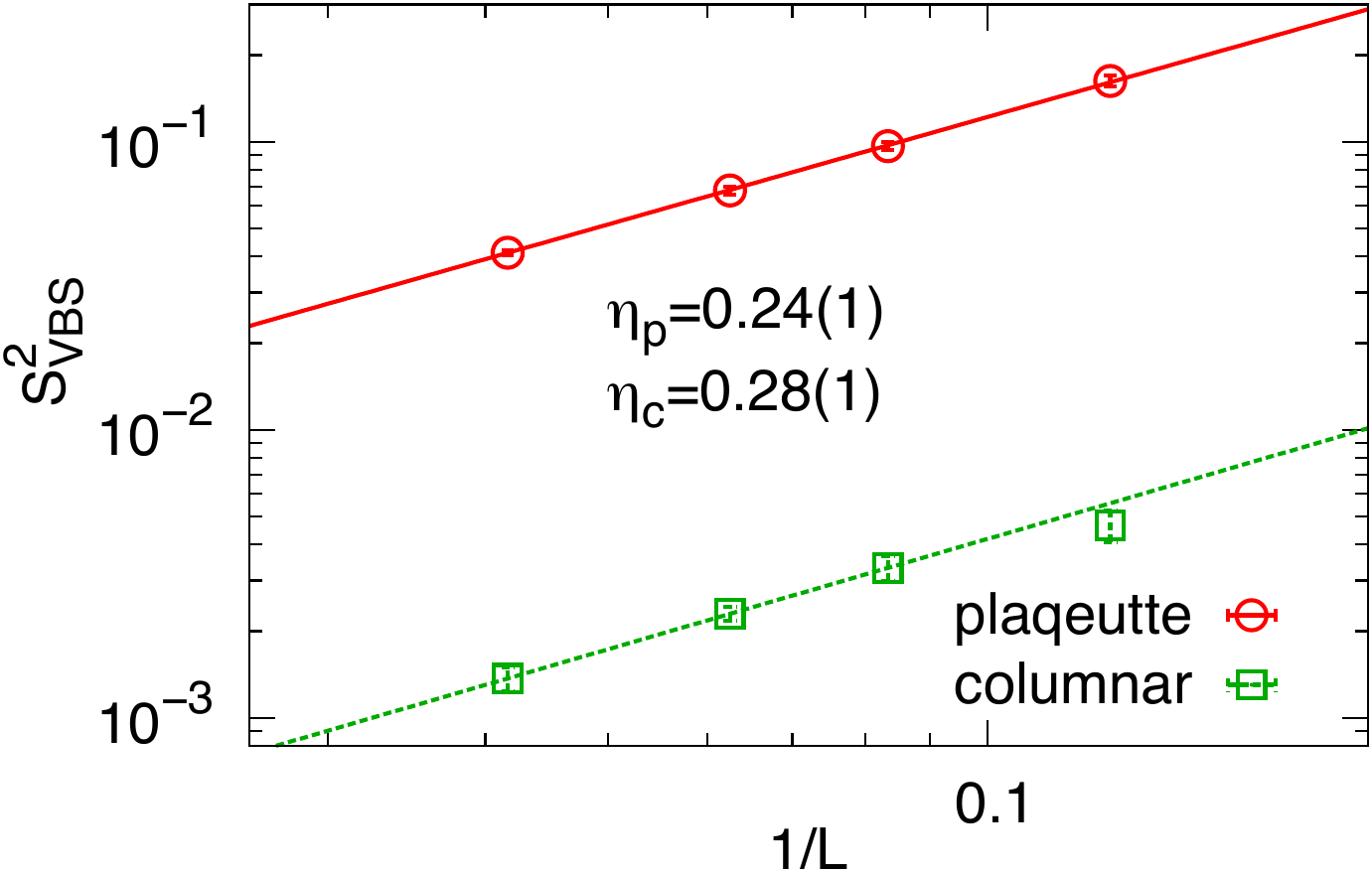}
\caption{The valence bond solid order parameters $S^2_{col}$ and $S^2_{plq}$ at $J_2=0.6$ as functions of $1/L$ in {\it log-log} plot. The power law decay behaviors are captured by decay exponents $1+\eta_p=1.24(1)$ and $1+\eta_c=1.28(1)$ for the plaquette and columnar VBS order respectively.}
\label{vbsord}
\end{center}
\end{figure}

\emph{Valence bond solid orders} -- To determine the phase at region $J_2\in
(0.572,0.6]$, dimer-dimer and plaquette-plaquette correlation functions
are investigated. We define dimer-dimer correlation function as
\begin{eqnarray}
C_{dx}(r_x,r_y)=\frac{1}{N}\sum_{x,y}D_{x}(x,y)D_{x}(x+r_x,y+r_y),\\
\label{dmr-corr}
C_{dy}(r_x,r_y)=\frac{1}{N}\sum_{x,y}D_{y}(x,y)D_{y}(x+r_x,y+r_y),
\end{eqnarray}
with
\begin{eqnarray}
D_{x}(x,y)&\equiv &\mathbf{S}_{(x,y)}\cdot\mathbf{S}_{(x+1,y)},\\
D_{y}(x,y)&\equiv &\mathbf{S}_{(x,y)}\cdot\mathbf{S}_{(x,y+1)}.
\end{eqnarray}
Similarly, the plaquette correlation function
as~\cite{mambrini_prb74.144422,valentin}
\begin{equation}
C_{plq}(r_x,r_y)=\frac{1}{N}\sum_{x,y}Q(x,y)Q(x+r_x,y+r_y),
\end{equation}
with $Q(x,y)\equiv \left(
  P_{\square}(x,y)+P^{-1}_{\square}(x,y)\right)$ defined as the
permutation operator that permutes 4 spins on a plaquette by 1 lattice
spacing.

To subtract the background expectation value, we take the modified
correlation functions as following
\begin{eqnarray}
C^*_{dx}(r,r)&=& C_{dx}(r,r)-C_{dx}(r-1,r-1),\\
C^*_{plq}(r,r)&=&C_{plq}(r,r)-C_{plq}(r-1,r-1).
\end{eqnarray}
We present the modified dimer-dimer and plaquette-plaquette
correlation functions for $J_2=0.6$ and $L=8,12,16,24$ in
Fig.~\ref{correlationfuncs}. We find clear power law decay behaviors
for both the dimer-dimer and plaquette-plaquette correlations. To
measure the two most possible VBS orders, namely the {\it columnar}
VBS order and the {\it plaquette} VBS order, we define the following
order parameters
\begin{eqnarray}
S^2_{plq}(L)=\frac{1}{L-2}\sum_{r=2}^{L}(-1)^rC_{plq}(r,r),\\
S^2_{col}(L)=\frac{1}{L-2}\sum_{r=2}^{L}(-1)^rC_{dx}(r,r).
\end{eqnarray}
We show the above two VBS order parameters at $J_2=0.6$ as a function
of $1/L$ in {\it log-log} plot in Fig.~\ref{vbsord}. Again, we find
power law decay behaviors for both the {\it columnar} and {\it
  plaquette} VBS orders. Taking the critical scaling behavior as
$S_{\text{VBS}}^2(L)\varpropto L^{-(z+\eta)}$, we find the anomalous
plaquette scaling exponent $\eta_p=0.24(1)$ and the anomalous columnar
scaling exponent $\eta_c=0.28(1)$. Our critical exponents $\eta_s$,
$\eta_{c}$, $\eta_{p}$, and $\nu$ are all consistent with the results
of the $J-Q$ model~\cite{DQCP3,DQCP4,DQCP6}.

\emph{Gapless spin liquid vs. deconfined quantum criticality} -- The
exponentially decaying spin-spin correlation and power law decaying
dimer-dimer correlation indicate that the paramagnetic phase has a
spin $S=1$ gap but has no $S=0$ gap. These properties are consistent
with the gapless $U(1)$ spin liquid state constructed by a single
variational parameter TPS ansatz~\cite{WangJ1J2}. However, since it is
well known that a $U(1)$ gapless spin liquid state is unstable due to
the confinement of $U(1)$ gauge field in $2+1$ dimensions, we argue
that a VBS order with exponentially small amplitude might eventually
develop at long wave length. Thus, our numerical results could imply a
Landau forbidden phase transition from N\'eel order to VBS order
described by the DQCP scenario.

\section{Conclusions}
In conclusion, we applied the cluster update algorithm for tensor
product states (TPSs) to study the frustrated spin 1/2 $J_1$-$J_2$
antiferromagnetic Heisenberg model on square lattice. Limited to a
cluster size $2\times 2$, a rather large bond dimension $D=9$ is
feasible. Through a finite $D_c$ scaling, our ground state energies at
finite sizes are in good agreement with the results from a state of
art exact diagonalization (ED) study~\cite{richter_ed}, a $SU(2)$
symmetric density matrix renormalization group (DMRG)
study~\cite{ShengJ1J2}, and a variational Monte Carlo (VMC)
study~\cite{HuJ1J2}. Applying finite size scaling (FSS) to the
spin-spin correlation function, we found the staggered magnetization
diminishes to zero at $J_2^{c_1}=0.572(5)$, suggesting a continuous
quantum phase transition. 
We further observed an exponentially decaying spin-spin correlations
while power law decaying dimer-dimer and plaquette-plaquette
correlations up to $24\times 24$ system size. All these evidences
point to the the emergence of gapless $U(1)$ spin-liquid state
consistent with a single variational parameter TPSs
ansatze~\cite{WangJ1J2}. Nevertheless, since the $U(1)$ spin liquid is
unstable due to the instanton effect, a VBS order with a small
amplitude could emerge in the thermodynamic limit. Remarkably, we
found the critical exponents $\nu=0.50(8)$ and $\eta_s=0.28(6)$,
$\eta_p=0.24(1)$, $\eta_c=0.28(1)$, which agree with the observed
critical exponents for deconfined quantum critical point (DQCP) in the
$J-Q$ model on square lattice~\cite{DQCP3,DQCP4,DQCP6}. Thus our
numerical results strongly indicate a Landau forbidden phase
transition from Neel order to VBS order at $J_2^{c_1}$.

\section{Method}
\subsection{The cluster update imaginary time evolution algorithm}
The following is an illustration of how to construct the evolution
operators for this Hamiltonian. We expand the evolution operator
$\hat{\mathbf{O}}\sim\mbox{exp}\{-\epsilon J_1(\mathbf{S}_1\cdot
\mathbf{S}_2+\mathbf{S}_2\cdot \mathbf{S}_3)-2\epsilon
J_2\mathbf{S}_1\cdot \mathbf{S}_3\}$ on 3 sites from the Trotter
decomposition of the partition function. By writing
$\mbox{exp}(-\epsilon J \mathbf{S}_i\cdot \mathbf{S}_j)$ as
\begin{equation}
  \prod_{\alpha}\left[\cosh\left(\epsilon
      J/4\right)\mathbb{1}_i\otimes \mathbb{1}_j-\sinh\left(\epsilon
      J/4\right)\sigma^{\alpha}_i\otimes\sigma^{\alpha}_j\right],
\end{equation}
where $\alpha=x,y,z$, $\sigma^{\alpha}$ are Pauli matrices, and
omitting higher orders of $O(\epsilon)$, one obtains
\begin{eqnarray}
\nonumber
\hat{\mathbf{O}}&=&\mathbb{1}_1\otimes\mathbb{1}_2\otimes\mathbb{1}_3-\sum_{\alpha}\tanh(\epsilon J_1/4)
\sigma_1^{\alpha}\otimes\sigma_2^{\alpha}\otimes\mathbb{1}_3\\
\nonumber
&&-\sum_{\alpha}\tanh(\epsilon J_1/4)\mathbb{1}_1\otimes\sigma_2^{\alpha}\otimes\sigma_3^{\alpha}\\
&&-\sum_{\alpha}\tanh(\epsilon
J_2/2)\sigma_1^{\alpha}\otimes\mathbb{1}_2\otimes\sigma_3^{\alpha}.
\label{operator}
\end{eqnarray}
 The above
terms can be expressed as a matrix product operator
(MPO)~\cite{bogdan},
\begin{eqnarray}
&&\hat{\mathbf{O}}=\sum_{i_1,i_2,i_3=0}^{3}(\mathbf{v}_{i_1}^{T}\mathbf{B}_{i_2}\mathbf{v}_{i_3})\mathbf{X}_{i_1}\otimes
\mathbf{X}_{i_2}\otimes\mathbf{X}_{i_3}\nonumber\\
&&\mathbf{X}_0=\mathbb{1},\quad\mathbf{X}_1=\sigma^{x},\quad\mathbf{X}_2=\sigma^{y},\quad\mathbf{X}_3=\sigma^{z},\nonumber\\
&&\mathbf{v}_0=|0\rangle,\nonumber\\
&&\mathbf{v}_i=a|i\rangle,\quad (i=1,2,3),\nonumber\\
&&\mathbf{B}_0=|0\rangle\langle 0|+b|1\rangle\langle 1| +b|2\rangle\langle 2| +b|3\rangle\langle 3|,\nonumber\\
&&\mathbf{B}_i=c|0\rangle\langle i|+c|i\rangle\langle 0|,\quad
(i=1,2,3),
\end{eqnarray}
where $\mathbf{v}_i$ are the vectors of length 4, $\mathbf{B}_i$ are
$4\times 4$ matrices, $\mathbf{X_i}$ are operators acting on the
physical index, and $a,b,c$ are scalar variables. In order to
correctly match the coefficients in front of each term in
Eq.~(\ref{operator}), $a,b,c$ have to be chosen to satisfy
$ac=-\tanh (\epsilon J_1/4)$, $a^2b=-\tanh (\epsilon J_2/2)$, and
$|a|,|b|,|c|\ll 1$. Thus the evolution operators on these sites are
written as
$\hat{\mathbf{O}}_1=\sum_i\mathbf{v}_i^{T}\otimes\mathbf{X}_i$,
$\hat{\mathbf{O}}_2=\sum_i\mathbf{B}_i\otimes\mathbf{X}_i$ and
$\hat{\mathbf{O}}_3=\sum_i\mathbf{v}_i\otimes\mathbf{X}_i$
respectively.

\begin{figure}
\begin{center}
\includegraphics[width=7cm]{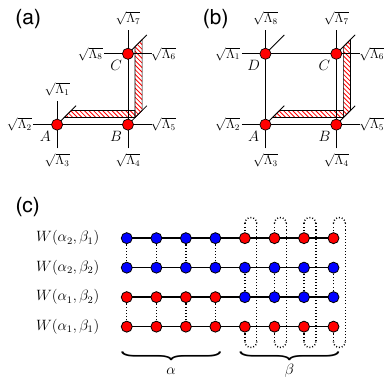}
\caption{(a) The simple update scheme. (b) The cluster update scheme
  with a cluster size $2\times 2$.}
\label{evolution1}
\end{center}
\end{figure}

We present the diagrammatic representation of the evolution
operators $\hat{\mathbf{O}}_1$, $\hat{\mathbf{O}}_2$ and
$\hat{\mathbf{O}}_3$ acting on sites $A$, $B$ and $C$ in a $2\times
2$ cluster in Fig.~\ref{evolution1}(b). The corresponding simple
update scheme is sketched in Fig.~\ref{evolution1}(a). In both
cases, the complexity scales as $D^5$, and there is no cumulative
error.

\section*{ Acknowledgments}
We would like to thank J.~Richter for passing their exact diagonalization
data for comparison, thank F.~Becca for passing their variational monte carlo
data for comparison; and thank A.~W.~Sandvik, Leon Balents, H.-C.~Jiang,
D.~Sheng, S.-S.~Gong and Z.-Y.~Zhu for their stimulating discussion. L.W. is
supported by National Natural Science Foundation of China (NSFC-11474016),
National Thousand Young Talents program of China. Z.C.G. is supported by
startup support from Department of Physics, CUHK. F.V. is supported by the EU
Strep project QUEVADIS, the ERC grant QUERG, and the FWF SFB grants FoQuS and
ViCoM. The computational results presented have been achieved partially using
Tianhe-2JK computing time award at the Beijing Computational Science Research
Center (CSRC).

\section*{Supplementary material}
\begin{widetext}
\begin{table}
\begin{tabular}{|l|l|l|l|l|l|l|l|l|l|}
  \hline\hline
 & L=4 &L=4 (ED)& L=6 & L=6 (ED) & L=8 & L=12 & L=16 & L=$\infty$ & ED($\infty$)\\\hline
$J_2=0.60$ & -0.5177(15) & -0.525896 & -0.4879(11) & -0.49323859 & -0.4808(9) & -0.4784(3) & -0.4777(3) & -0.4773(1) & -0.4811(82) \\\hline

$J_2=0.55$ & -0.5192(11) & -0.523595 & -0.4933(5) & -0.49517770 & -0.4880(2) & -0.4862(2) & -0.4859(4) &  -0.4857(2) & -0.4932(47) \\\hline

$J_2=0.52$ & -0.5227(5) & -0.525938 & -0.4983(6) & -0.49988464 & -0.4941(1) & -0.4920(1) & -0.4920(1) & -0.4916(1) & -0.4975(36) \\\hline

$J_2=0.50$ & -0.5268(4) & -0.52862 & -0.5034(3) & -0.50380965 & -0.4984(2) & -0.4963(2) & -0.4966(3) & -0.4958(3) & -0.5012(30) \\\hline

$J_2=0.48$ & -0.5307(6) & -0.532034 & -0.5072(5) & -0.50823276 & -0.5034(4) & -0.5008(1) & -0.5008(2) & -0.5004(2) & -0.5053(26) \\\hline

$J_2=0.45$ & -0.5370(4) & -0.5383 & -0.5151(1) & -0.51565739 & -0.51080(2) & -0.5088(1) & -0.5087(1) & -0.5084(1) & -0.5122(20) \\\hline

\hline
\end{tabular}
\caption{{\label{enrtab}}Finite size ground state energies and their extrapolation to the thermodynamic limit, compared with extrapolation from ED study.}
\end{table}
\end{widetext}

\end{document}